\documentclass[aps,prl,twocolumn,showpacs,superscriptaddress,groupedaddress]{revtex4-1}
\usepackage{graphicx}
\usepackage{amsmath}
\usepackage{color}
\usepackage{epstopdf}
\usepackage{dcolumn}   
\usepackage{bm}        
\usepackage{amssymb}  
\def\Vg{V_{G}}
\def\Eth{E_{Th}}
\def\Ic{I_{C}}
\def\Ir{I_{R}}
\def\Is{I_{S}}
\def\Rn{R_{N}}

\begin{document}

\title{Critical current scaling in long diffusive graphene-based Josephson junctions}
	
	\author{Chung-Ting Ke}
	\affiliation{Department of Physics, Duke University, USA}
	\author{Ivan V. Borzenets}
	\affiliation{Department of Applied Physics, The University of Tokyo, Japan}
	\author{Anne W. Draelos}
	\affiliation{Department of Physics, Duke University, USA}
	\author{Fran\c{c}ois Amet}
	\affiliation{Department of Physics, Duke University, USA}
	\author{Yuriy Bomze}
	\affiliation{Department of Physics, Duke University, USA}
	\author{Gareth Jones}
	\affiliation{College of Engineering, Mathematics and Physical Sciences, University of Exeter, UK }
	\author{Monica Craciun}
	\affiliation{College of Engineering, Mathematics and Physical Sciences, University of Exeter, UK }
	\author{Saverio Russo}
	\affiliation{College of Engineering, Mathematics and Physical Sciences, University of Exeter, UK }
	\author{Michihisa Yamamoto}
	\affiliation{Department of Applied Physics, The University of Tokyo, Japan}
	\author{Seigo Tarucha}
	\affiliation{Department of Applied Physics, The University of Tokyo, Japan}
	\affiliation{RIKEN, Center for Emergent Matter Science, Japan}
	\author{Gleb Finkelstein}
	\affiliation{Department of Physics, Duke University, USA}

	\def\GBJJ{graphene-based Josephson junction}
	\def\GBJJs{graphene-based Josephson junctions }
	\def\JJ{Josephson junction}

	\begin{abstract}		
	We present transport measurements on long diffusive graphene-based Josephson junctions. Several junctions are made on a single-domain crystal of CVD graphene and feature the same contact width of $\sim9\mu$m but vary in length from 400 to 1000 nm. As the carrier density is tuned with the gate voltage, the critical current in the these junctions spans a range from a few nA up to more than $5\mu$A, while the Thouless energy, $\Eth$, covers almost two orders of magnitude. Over much of this range, the product of the critical current and the normal resistance $\Ic \Rn$ is found to scale linearly with $\Eth$, as expected from theory. However, the ratio $\Ic \Rn /\Eth$ is found to be 0.1-0.2: much smaller than the predicted $\sim$10 for long diffusive SNS junctions.
\end{abstract}    
		\maketitle

Electrical current can flow without dissipation through a normal (non-superconducting) material connected to superconducting contacts~\cite{Tinkham}.
Using graphene as a normal region, one can create gate-tunable superconducting devices which feature both a high electronic mobility and a large Fermi velocity~\cite{heersche_2007,miao_2007,du_2008,gueron_2009}. As a result, in the cleanest devices supercurrents can propagate ballistically on a micron scale~\cite{Ballistic_1, Ballistic_2}. However, the mechanism allowing supercurrent transport through diffusive graphene is not yet fully understood. Even in relatively disordered samples, supercurrents can propagate through channel lengths exceeding one micron; however, the channel length dependence of the critical current has not yet been explored. 
		
		 Conventional theory for diffusive Superconductor-Normal Metal-Superconductor (SNS) junctions predicts that the critical current of the junction ($\Ic$) is determined by the Thouless energy $\Eth$$\,=\,$$\hbar D/L^{2}$, where $D$ is the diffusion coefficient and $L$ the channel length. For long junctions where $\Eth$$\,<<\,$$\Delta$, $\Ic$ is predicted to be given by $e\Ic \Rn \approx 10\Eth$~\cite{Dubos2001}. Here, $\Delta$ is the superconducting gap in the leads, and $\Rn $ is the normal state resistance of the junction. In this work, we investigate the relationship between $\Ic \Rn $ and $\Eth$ in Josephson junctions of different lengths made on the same graphene crystal. We establish that in a wide range of critical currents (covering more than two orders of magnitude) and densities away from the Dirac point, $\Ic \Rn $ is indeed proportional to $\Eth$. However, the coefficient of proportionality is significantly suppressed compared to the expected value of $\sim$10. 
	
Our devices utilize large domain size ($\sim100\mu$m) graphene grown via Chemical Vapor Deposition (CVD)~\cite{CVD_1,CVD_2,Kim2009,CVD_4,CVD_5}. A macroscopic piece of graphene film, grown on Copper foil~\cite{CVD_1}, is transferred to a 300 nm oxide SiO$_2$/Si substrate using the standard PMMA/FeCl$_3$ transfer technique~\cite{Kim2009}. The superconducting contacts are made by depositing 120 nm of lead (Pb) on top of a 6 nm Pd sticking layer, which provides an electrically transparent contact to graphene. Pb has a high critical temperature $T_{C}\sim7$ K, so a supercurrent can be observed through graphene junctions at temperatures up to 4 K~\cite{Borzenets2011, Borzenets2012}. Our three junctions have lengths (distance between the contacts) of $L=$ 400, 600 and 1000 nm respectively [Figure \ref{figure:Fig1}(a)]. The width of the contacts is $W=9\mu$m, which yields a sufficiently high supercurrent even for the longest one. The junctions are isolated from the large capacitances of the bonding pads by on-chip resistors in the range of 500 $\Omega$ to a few $k\Omega$ placed just outside the junctions. These resistors are created by partially oxidizing the Pb contact outside of the junctions using oxygen plasma. 
The CVD graphene is not encapsulated, suspended, or annealed; the resulting electron mean free path is less than $100$ nm, as estimated from the bulk resistance. Thus, the devices are definitely in the diffusive junction limit~\cite{Dubos2001}.
	
	\begin{figure}[htbp]
			\centering
			\includegraphics[width=0.5\textwidth]{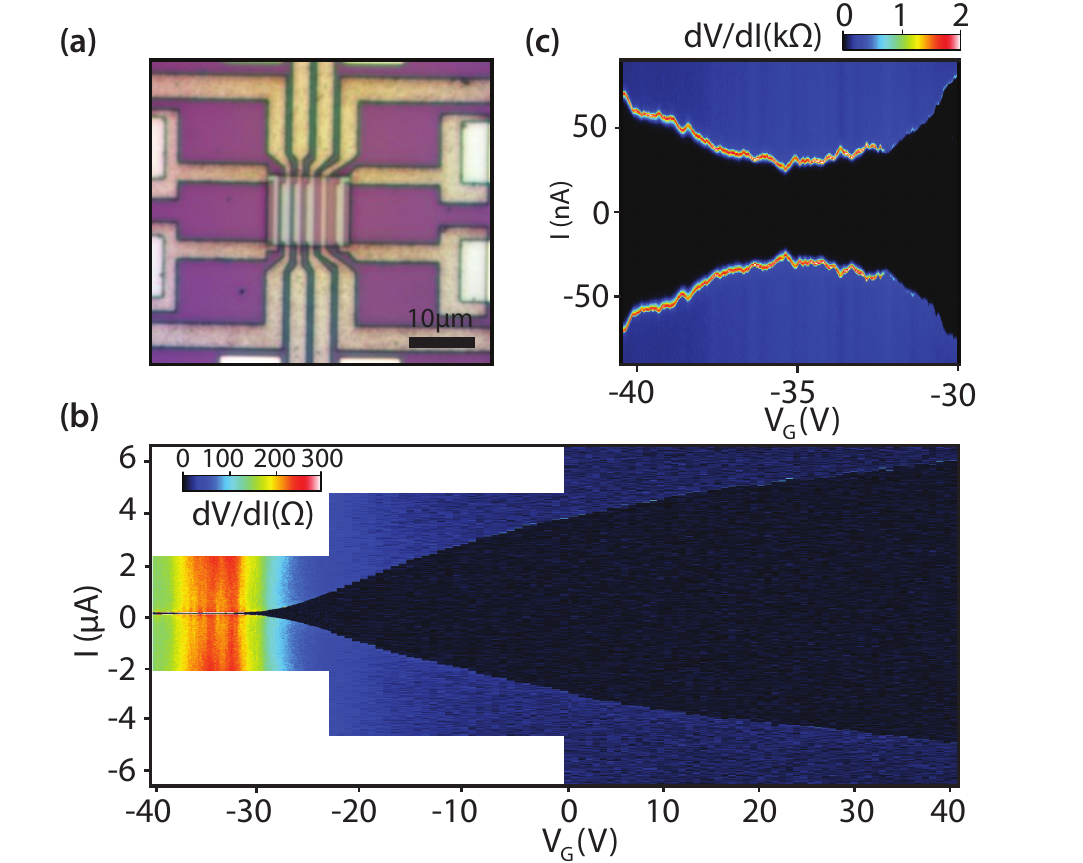}
			\caption{(a) Optical microscope image of graphene-based Josephson junction device. A $9 \times 20\mu$m active area is defined in CVD graphene and the unused material is etched away. Josephson junctions of the same width, but varying lengths ($400$ nm, 600 nm,  1000 nm, 1500 nm,  2000 nm) are made by depositing superconducting metal contacts onto the graphene. In order to form resistors that isolate the junctions from the rest of the circuit, parts of electrodes located away from the active area are oxidized with oxygen plasma. Only the first three junctions are used in this work. (b) Differential resistance $dV/dI$ measured versus applied current and gate voltage in the 400 nm-long junction. The full back gate range from $39$ V to $-39$ V is divided in three separate maps with different current ranges. The superconducting region is observed around zero current (dark region of vanishing resistance.) The current is swept from negative to positive bias. $\Ir$, the retrapping current from the normal to the superconducting state (observed at negative current) is smaller but comparable to $\Is$, the switching current from the superconducting to the normal state (at positive current.) (c) Zoomed-in resistance map around the Dirac point.   {\label{figure:Fig1}}}
	\end{figure} 
		
	The measurements are performed in a dilution refrigerator with a base temperature of 50 mK. The device is isolated via low temperature RC filters, resistive coaxial lines, as well as RF shielding. Figure \ref{figure:Fig1}(b) shows the differential resistance $dV/dI$ map measured versus the applied current $I$ and the back-gate voltage $\Vg$ for the $400$ nm device. The area of zero resistance (black region) indicates the superconducting regime, which persists for all values of the gate voltage. In this measurement, the current is swept from negative to positive; thus, at negative bias, the device transitions from the normal to the superconducting state, yielding the retrapping current $\Ir$. At positive current, the junction switches from the superconducting to the normal state at the switching current, $\Is$~\cite{Tinkham}. $\Is$ is found to be greater, but comparable to $\Ir$ throughout the gate voltage range. We have earlier attributed the origin of this hysteresis in similar resistively isolated graphene junctions to electron heating~\cite{PRL2013}. In the following, we use $\Is$ to represent the true critical current of the sample, $\Ic$.
	
	 Figure \ref{figure:Fig1}(c) shows a high resolution $dV/dI$ map measured around the Dirac point (DP), where the switching current $\Is$ is the lowest. The DP in this junction is located at $\Vg$$\,=\,-35$ V, and the DP in other junctions are located within 3 V of this value, indicating that our graphene is uniformly N-doped. In the following, we plot the voltage as measured from the Dirac point, $V_G-V_D$.
	
	\begin{figure}[htbp]	
		\centering
		\includegraphics[width=0.5\textwidth]{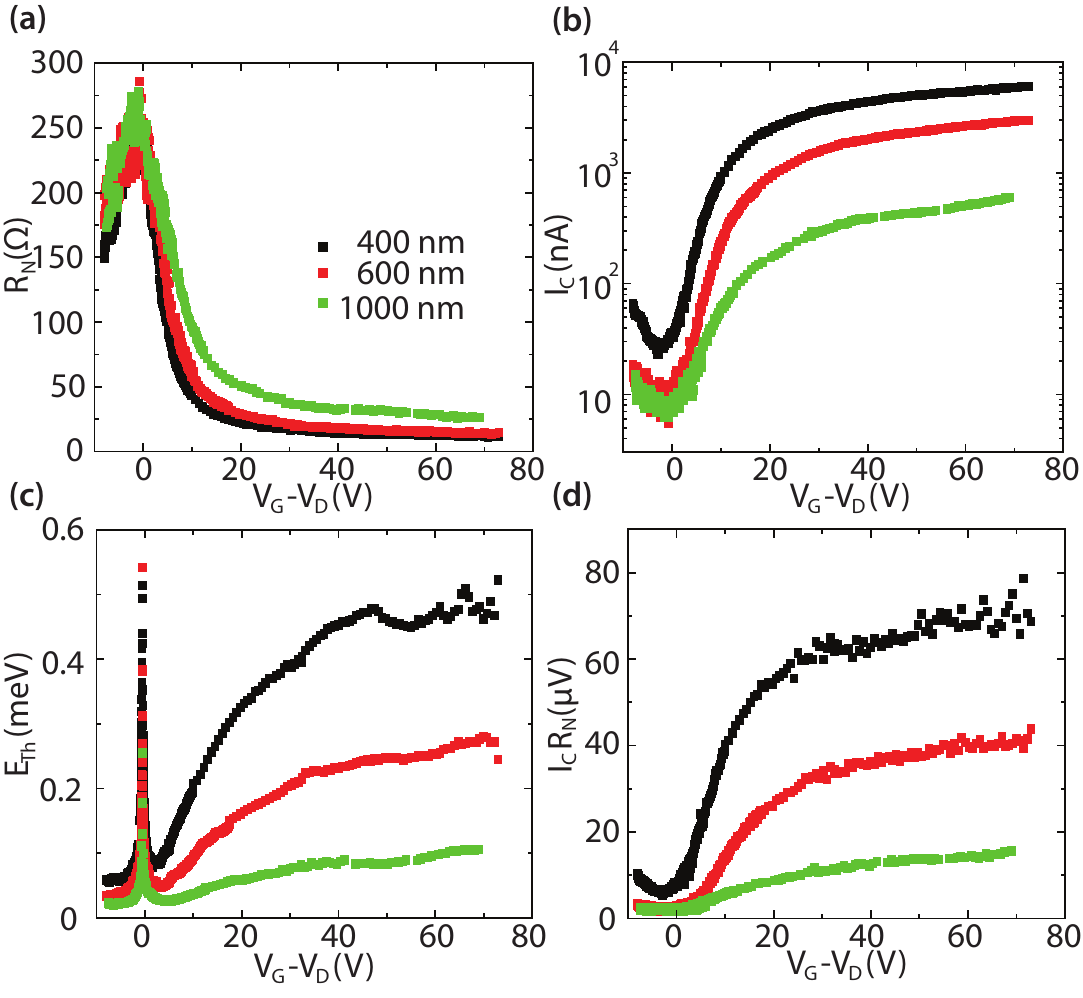}
		\caption{(a) Normal resistance, (b) critical current, (c) calculated Thouless energy, and (d) $I_C R_N$ product, plotted versus the gate voltage measured from the Dirac point, $\Vg-V_D$, for the three junctions of different lengths. In panel (c), the $\Eth$ curves artificially diverge at the DP, because the average density goes to zero, while the resistance stays finite. We therefore ignore any data taken with $V_G$ within 7 V from the DP. Outside of this regime, $\Eth$ is sufficiently (more than 5 times) larger than $k_B T$, which allows us to use the zero-temperature limit of Ref.~\cite{Dubos2001}. It is clear that the shape and the relative values of the three curves in panel (c) closely resemble those of panel (d), except in the vicinity of the DP. {\label{figure:Fig2}}}	
	\end{figure}
	
	Figure \ref{figure:Fig2}(a,b) shows the normal resistance $\Rn$ and the switching current $I_S$ versus $\Vg$ for all three junctions. Naturally, the switching current decreases with increasing channel length. The minimum $I_S$ measured at the Dirac point is $\sim25$$\,$nA for the $400$$\,$nm junction and drops to $\sim8$$\,$nA for the $1000$$\,$nm junction. Since these small values are strongly affected by fluctuations, we exclude them from the analysis. 

	In long diffusive SNS junctions, the product of the critical current and normal resistance $\Ic \Rn $ is expected to be proportional to $\Eth << \Delta$~\cite{Dubos2001}. The Thouless energy $\Eth$$\,=\,$$\hbar D/L^2$ can be further expressed as a function of conductivity $\sigma$, and the density of states ${\partial n}/{\partial \varepsilon}$. The conductivity could be obtained from $R_N$ as $\sigma$$\,=\,$$L/(R_N-R_C) W$, where $W$ is the width of the junction and $R_C$ is the contact resistance.  $R_C$  for each junction can be fitted by relying on the fact that $\sigma$ of diffusive graphene scales linearly with gate voltage measured from the Dirac point, so that $1/(R_N-R_C)$ should be $\propto(V_G-V_D)$~\cite{Geim_2009}. The resulting $R_C$ values for the three junctions are found to be in the range of $\approx$ 6-14 $\Omega$.  The Thouless energy may be eventually rewritten as:   
	\begin{equation}
		\Eth= \dfrac{\hbar\sigma}{e^2L^2}(\dfrac{\partial \varepsilon}{\partial n})\propto \frac{1}{(\Rn-R_C) L W \sqrt{n}} 
	\end{equation} \label{equ.1} where, $n$$\,\equiv\,$$C(\Vg-V_{D})/e$ is the carrier density and $C$ is the gate capacitance. 

Figure \ref{figure:Fig2}(c) shows the resulting dependence of the Thouless energy on $V_G$, as calculated from eq. (1). For a junction to be in the regime where $\Ic \Rn \propto \Eth$,  the Thouless energy must be much larger than $k_BT$. This condition holds for a wide range of densities in all three junctions since $k_BT$$\,\sim\,$4.3$\,\mu$eV for $T=50$ mK. Around the neutrality point, the carrier density is inhomogeneous and fluctuates strongly in space~\cite{Yacoby}. In this regime the calculation of $\Eth$ becomes invalid; the very low density regime is therefore excluded from future analysis.

	Finally, Figure \ref{figure:Fig2}(d) shows the product of $\Ic \Rn$ for the same junction. We expect the relation between the critical current and the Thouless energy to be of the form $e\Ic \Rn $$\,=\,$$\alpha \Eth$, and indeed the shapes of the curves in Figure \ref{figure:Fig2}(c,d) are similar except in the vicinity of the DP.  We further plot $\Ic$ as a function of the $\Eth/R_N$ ratio in Figure \ref{figure:Fig3}. We find that the data measured on the three junctions of different lengths over more than two orders of magnitude in $I_S$ collapse on the same linear curve, in agreement with the theory for diffusive SNS junctions~\cite{Dubos2001}.

	However, unlike the case of conventional metal SNS junctions ~\cite{Dubos2001,Dubos2001_2}, the ratio of $e\Ic \Rn / \Eth \sim$ 0.1-0.2 is suppressed by a factor of 50-100 compared to the expected value of $\alpha \approx 10$~\cite{Dubos2001}. The reduced value of the critical current in diffusive \GBJJs  has been observed in previous studies~\cite{Lee2011,Komatsu2012,Li}, and attributed to the partial transmission of the graphene/superconductor interface~\cite{Komatsu2012,Li}. Indeed, in case of the partial transmission, the diffusion time inside the junction, $\hbar/\Eth$ has to be lengthened by the inverse of the transparency $1/t$ of the superconductor-normal interface. As a result, $\Eth$ will be replaced by $\Eth^*$$\,=\,$$t\Eth$. 
To account for our observations, $t$ should be $\approx$ 0.01-0.02, independent of the junction and the electron density. This very low transmission likely occurs at the interface between the Pd contact layer and the highly doped graphene underneath. This interface has a very large number of modes which could have a small transmission, without contributing an overwhelmingly high resistance. Indeed, taking the density under the contact to be in the $10^{13} /cm^2$ range, and assuming that the effective coupling length between graphene and Pd is $\sim$100 nm, we estimate the number of transmission modes between graphene and metal to be in the $10^5$ range. Furthermore, unlike the bulk of graphene, the electron density in graphene under the contact is relatively insensitive to the gate voltage; therefore its interface with metal should produce a constant contribution to the resistance. Assuming $10^5$ modes with $t \approx 0.01$ transmission yields $\sim $10 $\Omega$, which is similar to the value of the contact resistance that we extracted earlier. 
	
	\begin{figure}
		\centering
		\includegraphics[width=0.4\textwidth]{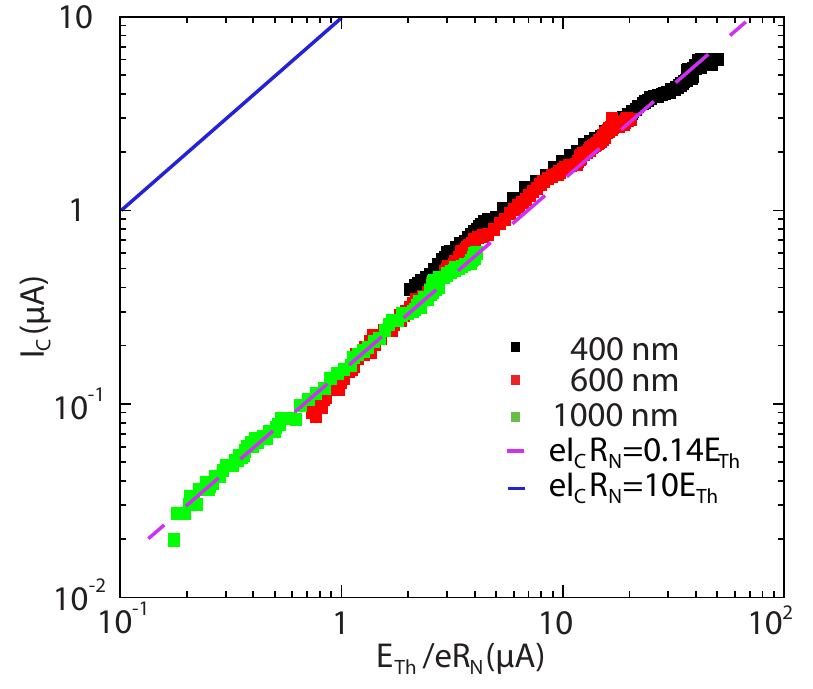}
		\caption{$I_C$ plotted vs. $\Eth / eR_N$ for the three junctions. Clearly, $I_C$ scales linearly with $\Eth / eR_N$ with the same coefficient for the three junctions. However, the proportionality coefficient is suppressed by a factor of 50-100 compared to the theoretical results of Ref.~\cite{Dubos2001}, as discussed in the text. The blue line indicates the expected scaling $eI_{C}R_{N}\approx10E_{Th}$ while the purple fit corresponds to $eI_{C}R_{N}$$\,=\,$$0.14E_{Th}$. {\label{figure:Fig3}}}			
	\end{figure}

	In summary, we studied the length and density dependence of the critical current in graphene-based Josephson junctions. The scaling function $e\Ic \Rn $$\,=\,$$\alpha \Eth$ works well away from the Dirac point, over a range of critical currents covering more than two orders of magnitude. However, we observe that $\alpha \sim 0.1-0.2$ , instead of $\sim 10$ as measured in metallic SNS junctions. This suppression may be attributed to the effective enhancement of the diffusion time due to suppressed transmission at the metal-graphene interfaces. 

\paragraph{Acknowledgments}

C.-T. K. and G.F. were supported by the Division of Materials Sciences and Engineering, Office of Basic Energy Sciences, U.S. Department of Energy, under Award No. DE-SC0002765. F.A. acknowledges support from the Fritz London postdoctoral fellowship and the ARO under Award W911NF-14-1-0349. I.V.B. and M.Y. are funded by the Canon foundation and Grants-in-Aid for Scientific Research on Innovative Areas, Science of Atomic Layers. S.T. acknowledges JSPS, Grant-in-Aid for Scientific Research S (26220710) and Project for Developing Innovation Systems of MEXT, Japan. S. R. and M. F. C. acknowledge financial support from EPSRC (Grant EP/J000396/1, EP/K017160, EP/K010050/1, EP/G036101/1, EP/M002438/1, EP/M001024/1), from the Royal Society Travel Exchange Grants 2012 and 2013 and from the Leverhulme Trust. A.W.D. acknowledges support from the National Science Foundation Graduate Research Fellowship Program (Grant DGF1106401).

\begin{figure}[h!]
	\centering
	\includegraphics[width=0.8\textwidth]{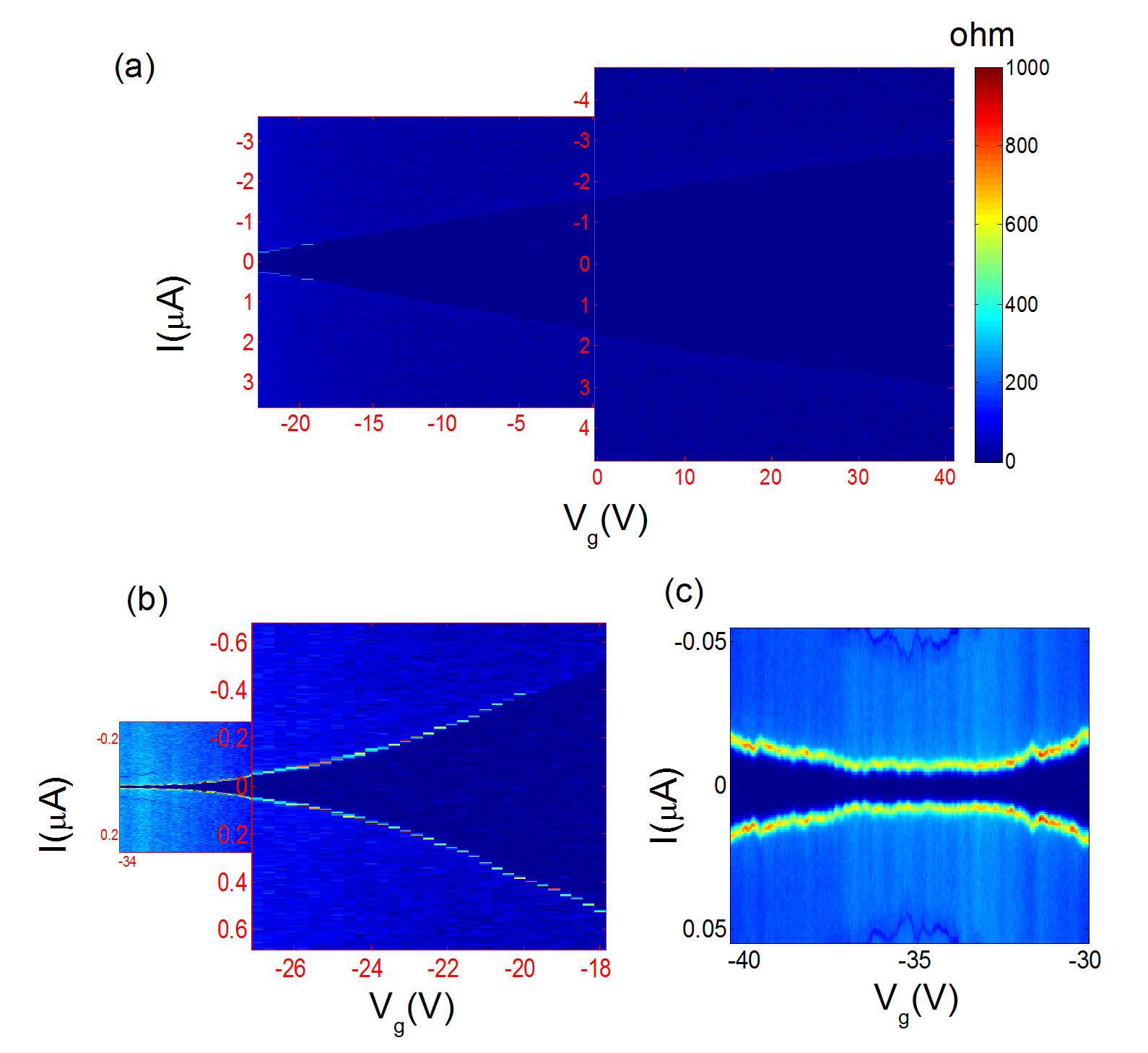}
	\caption{The current v.s. back gate dependence of the 600 nm \GBJJ. To see the detail around the DP, we divide the back gate into three regions. The tunneling current survives when the back gate is swept through the DP. The smallest critical current is found to be less than 10 nA. {\label{figure:Fig1}}}
\end{figure} 

\begin{figure}
	\centering
	\includegraphics[width=0.8\textwidth]{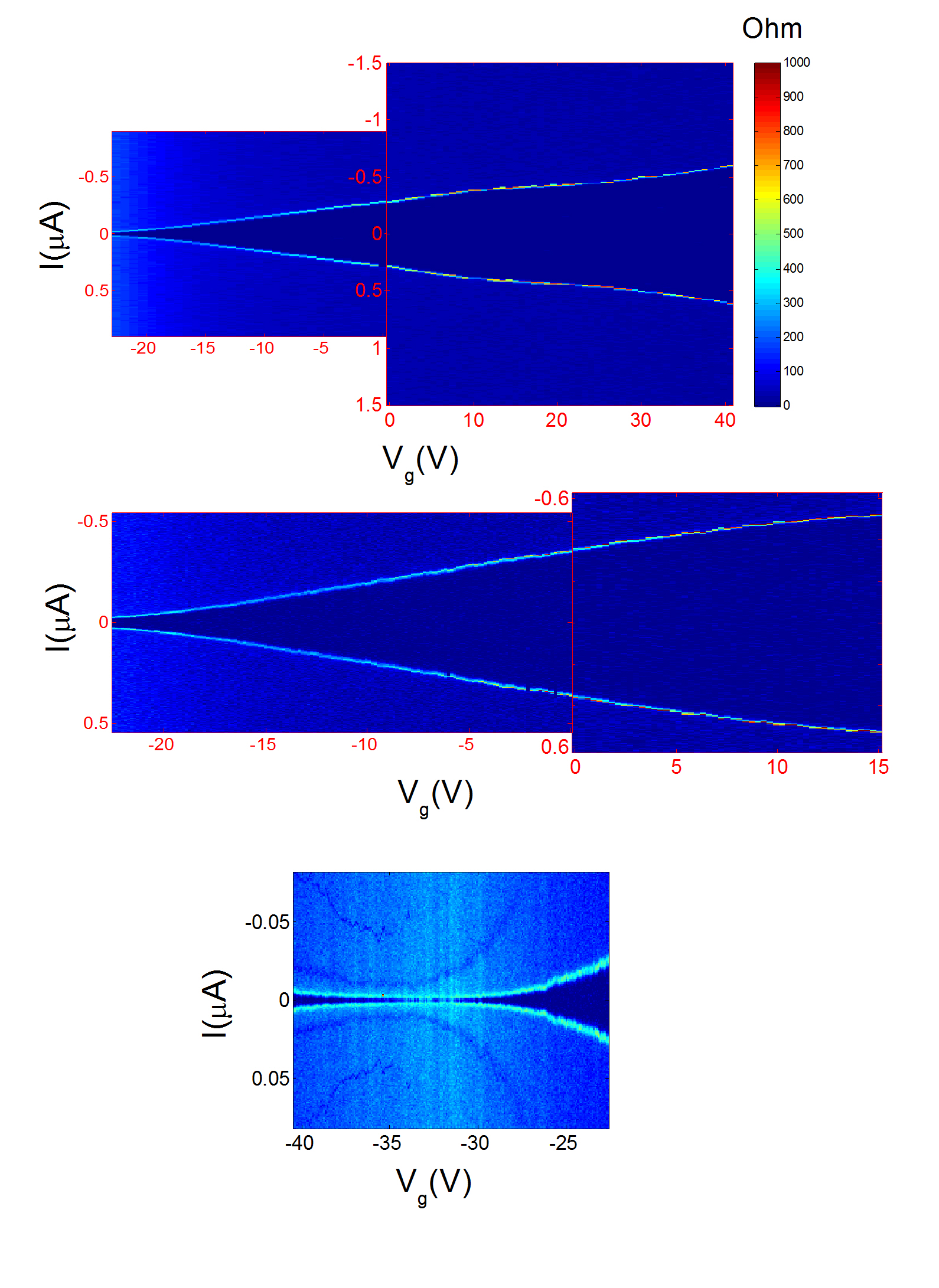}
		\caption{The current v.s. back gate dependence of the 1000 nm \GBJJ. We separate the current v.s. back gate into three different regimes. Approaching the DP, the current steps become smaller. In the vicinity of the DP, a very small but nonzero supercurrent can be observed.    {\label{figure:Fig1-1}}}
	
\end{figure} 




\begin{figure}
	\centering
	\includegraphics[width=0.8\textwidth]{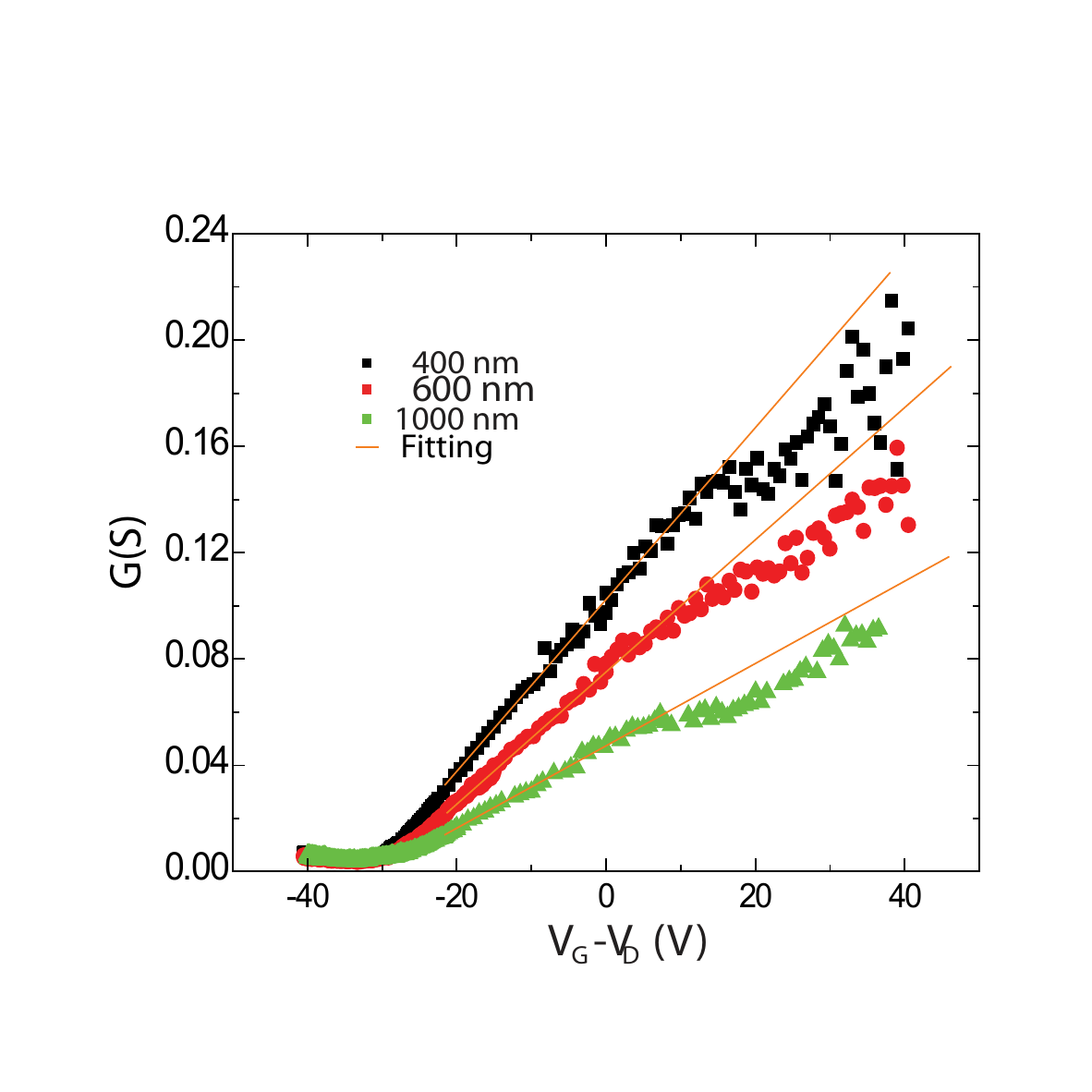}
	\caption{To fit the contact resistance for all three junctions, we plot the conductance v.s. the back gate dependence. In diffusive graphene, the conductance will scale linearly with back gate voltage away from the Dirac point. We start our fitting from -20V to 20 V to avoid the DP and high density noise regions.    {\label{figure:Fig3}}}
\end{figure} 

\begin{figure}
	\centering
	\includegraphics[width=0.8\textwidth]{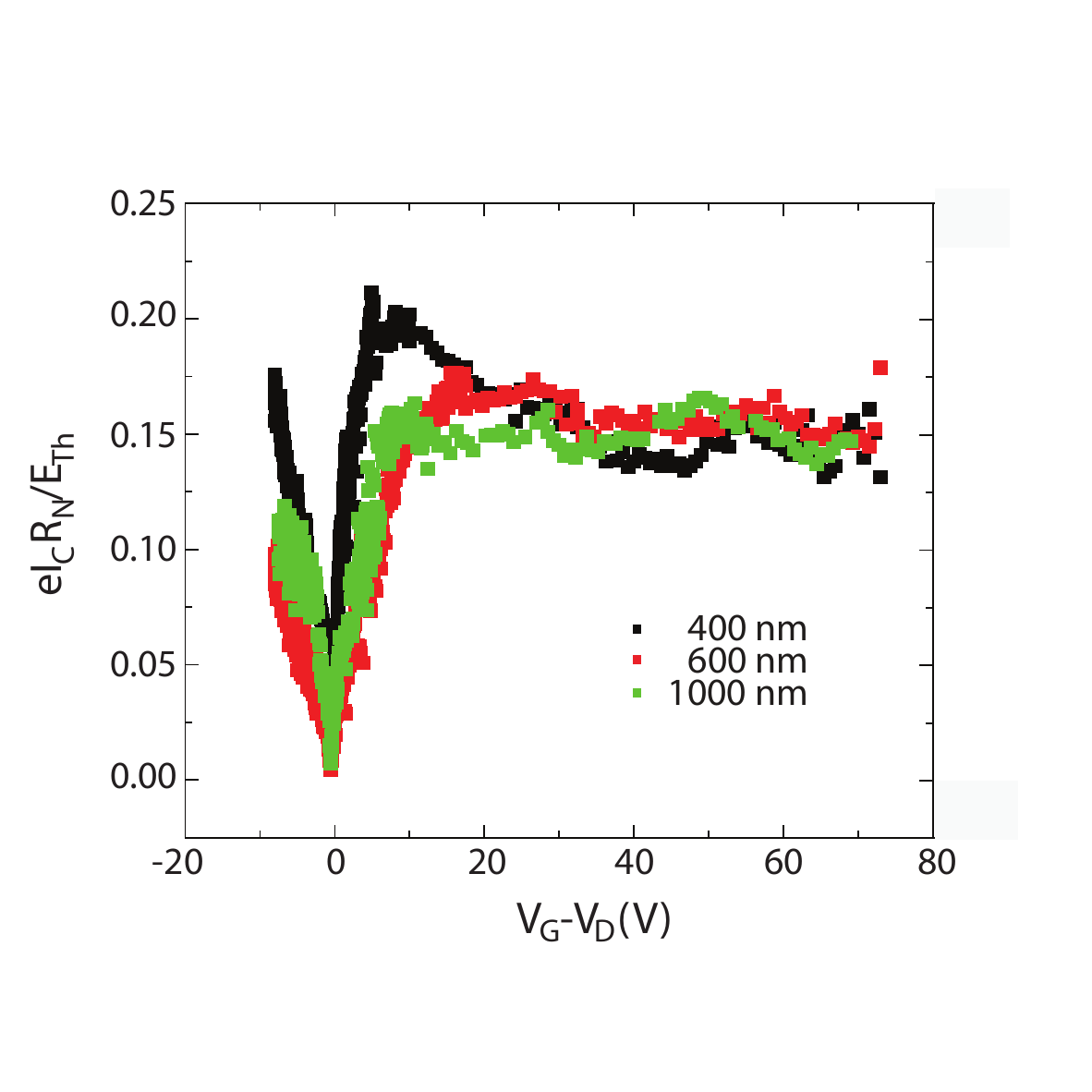}
	\caption{E$_{TH}$/I$_C$R$_N$ v.s. back gate for the \GBJJs.   {\label{figure:Fig4}}}
\end{figure}

\end{document}